# The 2D Power Spectrum of the Las Campanas Redshift Survey: Detection of Excess Power on 100 h$^{-1}$Mpc Scales

Stephen D. Landy and Stephen A. Shectman
Carnegie Observatories, 813 Santa Barbara Street, Pasadena, CA 94720

Huan Lin and Robert P. Kirshner
Harvard-Smithsonian Center for Astrophysics, 60 Garden Street, Cambridge, MA 02138

Augustus A. Oemler
Dept. of Astronomy, Yale University, New Haven, CT 06520-8101

and

Douglas Tucker
Astrophysikalisches Institut Potsdam, An der Sternwarte 16, D-14482 Potsdam, Germany

## ABSTRACT

We have measured the 2 dimensional (2D) power spectrum of the Las Campanas Redshift Survey on scales between 30 and 200 $h^{-1}$Mpc ($q_0 = 0.5$, $H_o = 100h$ km sec$^{-1}$ Mpc$^{-1}$). Such an analysis is more sensitive to structure on scales $> 50$ $h^{-1}$Mpc than a full 3 dimensional analysis given the geometry of the survey. We find a strong peak in the power spectrum at $\sim 100$ $h^{-1}$Mpc relative to the smooth continuum expected from the best fit Cold Dark Matter model ($Prob \sim 2.5 \times 10^{-4}$ with $\Omega h = 0.3$ assuming a Gaussian random field). This signal is detected in two independent directions on the sky and has been identified with numerous structures visible in the survey which appear as walls and voids. Therefore, we conclude that there exists a significant increase in power on this scale and that such structures are common features in the local universe, $z \leq 0.2$.

## 1. Introduction

The existence of large density fluctuations in the galaxy distribution on the order of 100 $h^{-1}$Mpc in size has been known since the discovery of the Boötes void in the early 1980's (Kirshner *et al.* 1981). Other early evidence of structure on such scales was found by Chincarini, Giovanelli, and Haynes (1983) and de Lapparent, Geller, and Huchra (1986). Subsequently Geller and Huchra (1989) reported the discovery of the 'Great Wall', a large wall-like distribution of galaxies over 100 $h^{-1}$Mpc in extent. More recently, Broadhurst *et al.* (1990) have claimed detection of a sharp spike in the 1D power spectrum around 128 $h^{-1}$Mpc using deep pencil-beam surveys.

Although it is not unexpected that large structures should exist according to currently popular working theories of large scale structure formation, such as a Cold-Dark Matter power spectrum coupled with the 'pancake' scenario of structure formation as predicted by Zel'dovich, the existence of excess power on a preferred scale is problematic. However, there do exist physically motivated reasons to expect some perturbation in the power spectrum on these scales. For example, this scale is on the order of the horizon size at mass-radiation equality, and an 'acoustic peak' in the power spectrum on these scales is a prediction of many standard cosmologies (see Peebles Sec 25 1993).

To make a definitive measurement of galaxy density fluctuations in this regime requires a large number of independent volumes situated over a broad region of the sky. The Las Campanas Redshift Survey includes over 26,000 galaxies with a mean redshift of $z = 0.1$ and is well-suited for making such measurements. The survey consists of six slices each approximately 1.5° thick in declination by 80° wide in right ascension. In each hemisphere the slices are centered on the same right ascension while being offset in declination by 3 or 6 degrees. The three slices in the South galactic hemisphere are located at -39°,-42°, and -45° Dec centered on $0^h45^m$ RA, and the three in the North at -3°,-6°, and -12° Dec centered on $12^h45^m$ RA. A more detailed description of the survey including the preparation of the redshift sample and various selection effects are discussed in detail in Shectman *et al.* (1995) and Lin *et al.* (1995a).

## 2. Power Spectrum Analysis

In order to estimate the 2D power spectrum we principally follow the analysis outlined by Peacock & Nicholson (1991) and Feldman, Kaiser, & Peacock (1994) with some minor changes (see also Burbridge & O'Dell 1972, Webster 1976, Park *et al.* 1994). Each slice was analysed separately. The coordinates of each slice were rotated in order to maximize the projection of the slice in the x-y plane. The data was cut to include only those galaxies with 10,000 km sec$^{-1}$ $\leq v_z \leq$ 45,000 km sec$^{-1}$, in order to minimize the effects of uncertainties in the selection function. For fields in which the number of redshifts observed was less than that observable given the magnitude limits for the field, the observed points were weighted by $(f_{obs})^{-1}$ where $f_{obs}$ is the observed fraction (see also Lin *et al.* 1995a). This weighting approximates uniform weighting for a magnitude limited sample.

The galaxies were expanded in plane waves and normalized in the standard manner. The Fourier expansion is given by

$$\mathbf{a_k} \equiv N_{wt}^{-1} \sum_i (f_{obs_i})^{-1} \exp\left(i\mathbf{k}\mathbf{x_i}\right) \tag{1}$$

where $N_{wt}$ is the weighted number of galaxies in the sample, $f_{obs_i}$ is the appropriate fraction, $\mathbf{x_i}$

is the galaxy's 2D coordinates $(x_i, y_i)$, and $\mathbf{a_k} = (a_{kx}, a_{ky})$ is the complex Fourier coefficient of the 2D Fourier transform at wavenumber $\mathbf{k} = (k_x, k_y)$. The complex Fourier coefficients $\mathbf{w_k}$ of the window function with power spectrum $W_{2D}(\mathbf{k})$ were determined using the same angular and radial selection function as that of the data but with fifty times the number of galaxies distributed randomly. The window function is simply the Fourier transform of the selection function. The Fourier transform of the random catalog was then subtracted from that of the data and the power spectrum calculated (see Feldman, Kaiser, & Peacock 1994).

Converting sums to integrals, the normalized convolved estimate of the 2D power spectrum $\hat{P}_{2D}(\mathbf{k})$ is:

$$\hat{P}_{2D}(\mathbf{k}) \equiv \frac{\int d^2\acute{\mathbf{k}}\ W_{2D}(\mathbf{k}-\acute{\mathbf{k}})\ \left[P_{2D}(\acute{\mathbf{k}}) + S_{2D}\right]}{\int d^2\mathbf{k}\ W_{2D}(\mathbf{k})}. \quad (2)$$

This expression shows the well-known result that the true power spectrum $P_{2D}(\mathbf{k})$ plus the shot noise $S_{2D}$ is convolved with the power spectrum of the 2D window function $W_{2D}(\mathbf{k})$ to give the measured signal $\hat{P}_{2D}(\mathbf{k})$. At the wavelengths reported in this *Letter*, the correction term due to subtraction of the DC level is negligible and will be ignored (Peacock & Nicholson 1991 eqn. 25).

### 2.1. Calculation of the 2D Power Spectrum

The 2D power spectrum $P_{2D}(k_x, k_y)$ is a projection of the 3D spectrum $P_{3D}(k_x, k_y, k_z)$ onto the $k_x - k_y$ plane by way of a window and depends on the geometry and orientation of the survey. As a heuristic, in this development one can approximate the survey as having an effective volume with dimensions $(d_x, d_y, d_z)$ where $d_z \ll d_x, d_y$, and model the window functions in each of these directions as Gaussians with width $\sigma_i \sim \pi/d$. In such a case the window function is separable and the relationship between the 2D and 3D power spectrum is easily seen. Calculations for a 1D projection and further discussion can be found in Szalay *et al.* (1991).

In a full 3D analysis neglecting the shot-noise, the power spectrum estimator is

$$\hat{P}_{3D}(k_x, k_y, k_z) = \frac{1}{(2\pi)^{\frac{3}{2}}\sigma_x\sigma_y\sigma_z} \int\int\int_{-\infty}^{\infty} d\acute{k}_x\ d\acute{k}_y\ d\acute{k}_z\ P_{3D}(\acute{k}_x, \acute{k}_y, \acute{k}_z)\ e^{-\frac{1}{2}(\frac{k_x-\acute{k}_x}{\sigma_x})^2}\ e^{-\frac{1}{2}(\frac{k_y-\acute{k}_y}{\sigma_y})^2}\ e^{-\frac{1}{2}(\frac{k_z-\acute{k}_z}{\sigma_z})^2}. \quad (3)$$

In our 2D analysis, the volume is collapsed along the z direction resulting in a planar survey with dimensions $(d_x, d_y)$ and a projection of power onto the $k_x - k_y$ plane. It can be shown that the resulting 2D spectrum is given by the $k_z = 0$ component of the fully convolved 3D power spectrum. Therefore, we may define

$$P_{2D}(\acute{k}_x, \acute{k}_y) \equiv \frac{1}{\sqrt{2\pi}\sigma_z} \int_{-\infty}^{\infty} d\acute{k}_z\ P_{3D}(\acute{k}_x, \acute{k}_y, \acute{k}_z)\ e^{-\frac{1}{2}(\frac{\acute{k}_z}{\sigma_z})^2} \quad (4)$$

which formally illustrates the projection. Then

$$\hat{P}_{2D}(k_x, k_y) = \frac{1}{2\pi\sigma_x\sigma_y} \int\int_{-\infty}^{\infty} dk_x^{'} \; dk_y^{'} \; P_{2D}(k_x^{'}, k_y^{'}) \; e^{-\frac{1}{2}(\frac{k_x - k_x^{'}}{\sigma_x})^2} \; e^{-\frac{1}{2}(\frac{k_y - k_y^{'}}{\sigma_y})^2} \tag{5}$$

for consistency with Equation 2. Note that our definition of the 2D spectrum includes the normalization factor $(\sqrt{2\pi}\sigma_z)^{-1}$ for direct comparision with the $\hat{P}_{3D}(k_x, k_y, k_z = 0)$ component of the full 3D spectrum. Therefore, our results are reported in units of $(h^{-1}\mathrm{Mpc})^3$. For all subsequent analysis the exact window functions have been calculated numerically and the Gaussian approximation has not been made.

## 3. Results

### 3.1. CDM Simulations

In order to check our results and investigate the effects of the survey's geometry, selection functions, and those due to redshift distortions, we performed an identical analysis on synthetic data kindly supplied by Changbom Park (see Park *et al.* 1994). Fifty-six realizations of sets of the three southern slices were culled from a large CDM simulation, $(576\ h^{-1}\mathrm{Mpc})^3$ with $\Omega h = 0.2$, using the same selection function and geometry as the survey's. The 2D power spectrum for each slice in a set was calculated and then averaged together to generate a mean power spectrum for each set. The mean result for all 56 sets is shown in Figure 1b together with error bars derived from this data. For all slices, the shot-noise power was less than 250 $(h^{-1}\mathrm{Mpc})^3$.

Also shown in Figure 1b is a line indicating the expected 2D power spectrum after projection and convolution of the underlying 3D power spectrum of the simulation. The window functions for each slice were calculated numerically and then used to project and convolve the analytic 3D spectrum. As may be seen, there is excellent agreement between the synthetic data and the expected analytic result. This indicates that effects due to wavelength dependent redshift distortions are expected to be minimal in this regime.

Also plotted is the envelope of one-sigma fluctuations based on the amplitude of the expected signal and the degrees of freedom calculated from the number of independent modes for a given wavenumber $|\mathbf{k}|$ given the effective area of the 2D window function (FWHM: $2.20 \times 10^{-4} h^{-2}\mathrm{Mpc}^2$). The good agreement shows that a formal analysis based on a model spectrum, calculated degrees of freedom, and Gaussian random field is highly accurate. A similar technique will by necessity be used with the survey data.

### 3.2. Las Campanas Survey Results

The mean result for all six survey slices is shown in Figure 1a. Figure 2 shows the individual measured power spectrum for each slice and the mean north and south signals. Since for the survey data it is not possible to know the underlying 3D power spectrum *a priori*, we calculated the expected power spectrum given a class of linear CDM models with $0.2 \leq \Omega h \leq 0.5$ (Efstathiou *et al.* 1992, Bond and Efstathiou 1984). These models were projected and convolved as above and fit to the measured signal between 30 $h^{-1}$Mpc and 70 $h^{-1}$Mpc using only their overall amplitude as a free parameter. The best fit spectrum was given by $\Omega h = 0.3$ and is shown in Figure 1a.

As is evident, the measured signal below 70 $h^{-1}$Mpc and the fit are in good agreement, however a strong excess of power appears at $\sim$ 100 $h^{-1}$Mpc. This peak is essentially unresolved since its width is approximately that expected solely from convolution with the window function. The error bars are derived from analysis of the degrees of freedom at each wavenumber. Significant excess power at wavelengths at or above 100 $h^{-1}$Mpc is evident in 5 out of the 6 slices in Figure 2.

To be conservative in our analysis of the significance of this peak, we consider each set of three slices as one independent sample. The mean power spectrum at this wavenumber, $|\mathbf{k}| = 0.067$, consists of an average over approximately 15 independent modes in the north and 15 in the south with 2 degrees of freedom in each mode giving a total of 60 degrees of freedom. The amplitude of the measured peak at this point is approximately 1.76 times the expected signal as given by the $\Omega h = 0.3$ model. Taking the null hypothesis that the local universe on these scales is drawn from a Gaussian random field (as reported by Feldman, Kaiser, & Peacock 1994), the significance of this peak is $\sim 2.5 \times 10^{-4}$, that is, a peak of this amplitude at this wavenumber would be expected by chance once in ~3900 similar surveys. Also shown for comparison is the best fit for an $\Omega h = 0.24$ model which was the result found for the updated one-in-six QDOT IRAS survey (Feldman, Kaiser, & Peacock 1994). In this case, the significance is $\sim 1.1 \times 10^{-3}$.

Amendola (1994) has shown that the significance of a peak may be overestimated given the null hypothesis of a Gaussian random field due to the higher order correlations in the density field, albeit, the simulations indicate that this effect is not important at these wavenumbers (Figure 1b). The significance may also be modified by considering that one is also able to measure peaks at other wavenumbers. However, the existence of a peak on these scales has been anticipated by Broadhurst *et al.* (1990) using 1D pencil-beam surveys.

## 4. Discussion and Analysis

### 4.1. Comparison to a 3D Analysis and Other Surveys

A question is what are the advantages of a 2D versus 3D analysis (see Lin *et al.* 1995b) at these wavenumbers. This is answered by comparing the differences between a 1D projection and 2D convolution versus a straight 3D convolution given the geometry of the survey. To first order the effective dimensions of the survey given a full set of three slices in one hemisphere are $300 \times 400 \times 50$ $h^{-1}$Mpc. In the geometry of the 2D analysis, the survey dimensions are $300 \times 400$ $h^{-1}$Mpc. This affords several significant advantages, the most obvious being that a 2D analysis can fairly sample wavelengths up to several hundred $h^{-1}$Mpc. On the other hand, a full 3D analysis samples wavevectors at all orientations to the planes of the slices. As a result, the measured signal becomes dominated by aliased power at wavelengths above ∼50 $h^{-1}$Mpc, the effective thickness of a set of slices.

A well-sampled 2D analysis also has advantages in regard to the detection of non-Gaussian structures since such structures show up as strong, localized peaks at a specific wavevector in the power spectrum. Depending on how the survey has cut through such structures, these peaks may show up at their true frequency or be projected to other wavelengths. However, the excess power will still show up as a strong, local peak and be easily detected in a 2D analysis. In a 3D analysis in this geometry, such peaks would be smoothed over many directions and wavelengths due to the large width of the window function in the third dimension making such sharp features more difficult to detect.

Other surveys, such as the CfA2 (Geller & Huchra 1989, Vogeley *et al.* 1992) and SSRS2 (da Costa *et al.* 1994), although sampling structure at higher densities, are limited at these wavelengths. Recent work by da Costa *et al.* (1994), report analysis of volumes with a depth of ∼130 $h^{-1}$Mpc. As these surveys contain only a few independent modes at this scale, it is not surprizing that they might not detect such a peak. This signal has also not been detected in recent analyses of the IRAS survey (see Fisher *et al.* 1993, Feldman, Kaiser, & Peacock 1994). The latter analysis has on the order of the same effective volume as ours, however the sampling density is a factor of 10 to 15 less and the shot noise is on the order of the signal. In our survey, the sampling density and signal-to-noise is a factor of ten greater. Also, it has been shown that the detection of structure is a function of sampling density and signal-to-noise (see Szapudi & Szalay 1995).

### 4.2. Identification of Contributing Structures in the Survey

In order to determine what structures are responsible for these spikes, maxima of the plane waves corresponding to the largest peaks above ∼ 100 $h^{-1}$Mpc have been overlaid on maps of the real space distribution of galaxies for a slice from each hemisphere, -12° and -39°. The appropriate peaks have been identified as well on the conjugate 2D power spectrum contour maps of the same slices. These are shown in Figure 3 and 4 included as Plates. The maxima of the plane waves, with phases from the Fourier transform, have been plotted as straight lines in these Figures. Large structures with the appearence of walls and voids being traced out by superclusters are being detected by the 2D power spectrum analysis, giving strong visual confirmation of the results.

It is not unexpected that the strongest peak on these plates correspond to a wavevector pointing in the redshift direction as this may be due to amplitude enhancements from redshift distortions (see Kaiser 1987). However, the existence of numerous other peaks and structures with different orientations at this same scale precludes interpretations based principally upon this effect.

### 4.3. Gaussian or Non-Gaussian Structures

Of interest to the theory of structure formation is whether this excess power is the result of the existence of non-Gaussian structures or is rather an inherent increase in power. Unfortunately power spectrum statistics are not that robust in making such a determination (see Amendola 1994). An analysis based upon the distribution of power spectrum amplitudes (see Szalay 1991) have proved inconclusive primarily due to the degree of smoothing from the window function. On the other hand, the appearence of the structures identified by the peaks is striking as is the partial coherence between the peaks as can be seen in Plates 1 and 2. Such coherence would not be expected in the case of a Gaussian random field.

## 5. Conclusions

We have detected multiple structures in the galaxy distribution in two independent regions on the sky which correspond to a peak in the 2D power spectrum on the order of 100 $h^{-1}$Mpc, a wavelength similar to that reported by Broadhurst *et al.* (1990). The survey contains approximately 30 independent modes on this scale. The probability of detecting such a signal by chance, with the assumptions of a Gaussian random field and the best fit linear CDM power spectrum is $2.5 \times 10^{-4}$ ($\Omega h = 0.3$ fit between 30 $h^{-1}$Mpc and 70 $h^{-1}$Mpc). The structures responsible for this signal have been identified and have the appearance of walls and voids. Therefore, such large structures appear to be common features of the local universe.

## 6. Acknowledgements

The authors would like to thank Anand Sivaramakrishnan and the referee Michael Vogeley for useful discussions and comments. We would especially like to thank Alexander Szalay for many productive insights into the analysis and understanding of this problem. We acknowledge principal support from NSF grant AST9220460.

---

This preprint was prepared with the AAS LaTeX macros v4.0.

Fig. 1.— a) The 2D power spectrum of the Las Campanas Redshift Survey and linear CDM power spectrum null hypothesis. A best fit power spectrum was derived by projecting and convolving linear CDM power spectra with $\Omega h$ between 0.2 and 0.5 and fitting the result to the data between 30 and 70 $h^{-1}$Mpc. The best fit was found to be $\Omega h = 0.3$ and is shown in the graph. The error bars are calculated using the expected signal together with degree of freedom analysis. Also plotted is the best fit spectrum with $\Omega h = 0.24$. This is shown for comparison with the findings of the one-in-six QDOT IRAS survey. b) The average signal from 56 sets of synthetic data with the same geometry and selection function as the 3 Southern slices (-39°,-42°, and -45°Dec). For the synthetic data the one sigma error bars are derived from the sets themselves. The one sigma error envelope is based on the calculated degrees of freedom at each wavelength together with the mean signal. The analytic spectrum is calculated using the underlying power spectrum in the synthetic data, $\Omega h = 0.2$, projected and convolved with the window functions. Note that the power spectrum amplitude of the simulation is significantly higher than that of the data.

Fig. 2.— Individual graphs showing the signals from each of the six slices together with the mean signals for the North (-3°,-6°, and -12°Dec) and South (-39°,-42°, and -45°Dec) sets. Both directions show significant excess power at $\sim$ 100 $h^{-1}$Mpc. Peaks at or above 100 $h^{-1}$Mpc are evident in 5 out of the 6 slices. One sigma error bars are shown for only the North and South means as the individual slices in each hemisphere are not independent. For all slices, the shot-noise power was less than 250 $(h^{-1}\text{Mpc})^3$

Fig. 3.— This Figure shows the 2D power spectrum as a function of frequency for the -12°Dec slice in the North. The power spectrum has been normalized to the expected $\Omega h = 0.3$ spectrum as a function of wavelength in order that fluctuations have the same expectation, (1.0), at all wavelengths. A signal of 6.1 corresponds to the expectation of only one peak of greater height over the entire graph. The width of the peaks in this Figure gives a rough idea of the size of an independent volume. The coordinate of each peak $(f_x, f_y)$ in the power spectrum corresponds to the normal of a plane wave in real space. The three highest peaks above 100 $h^{-1}$Mpc have been numbered and correspond to the graphs showing the real space distribution of galaxies. The straight lines indicate the maxima of these plane waves with the appropriate phase and noted wavelength. In this way it is easy to see which structures in real space are responsible for these peaks. Notice the multiple in phase overdensities in the real space maps.

Fig. 4.— This Figure shows the 2D power spectrum as a function of frequency for the -39°Dec slice in the South. The power spectrum has been normalized to the expected $\Omega h = 0.3$ spectrum as a function of wavelength in order that fluctuations have the same expectation, (1.0), at all wavelengths. A signal of 6.1 corresponds to the expectation of only one peak of greater height over the entire graph. The width of the peaks in this Figure gives a rough idea of the size of an independent volume. The coordinate of each peak $(f_x, f_y)$ in the power spectrum corresponds to the normal of a plane wave in real space. The three highest peaks above 100 $h^{-1}$Mpc have been numbered and correspond to the graphs showing the real space distribution of galaxies. The straight

lines indicate the maxima of these plane waves with the appropriate phase and noted wavelength. In this way it is possible to see which structures in real space are responsible for these peaks. In this slice, the structures detected are not as well defined as those in Figure 3.

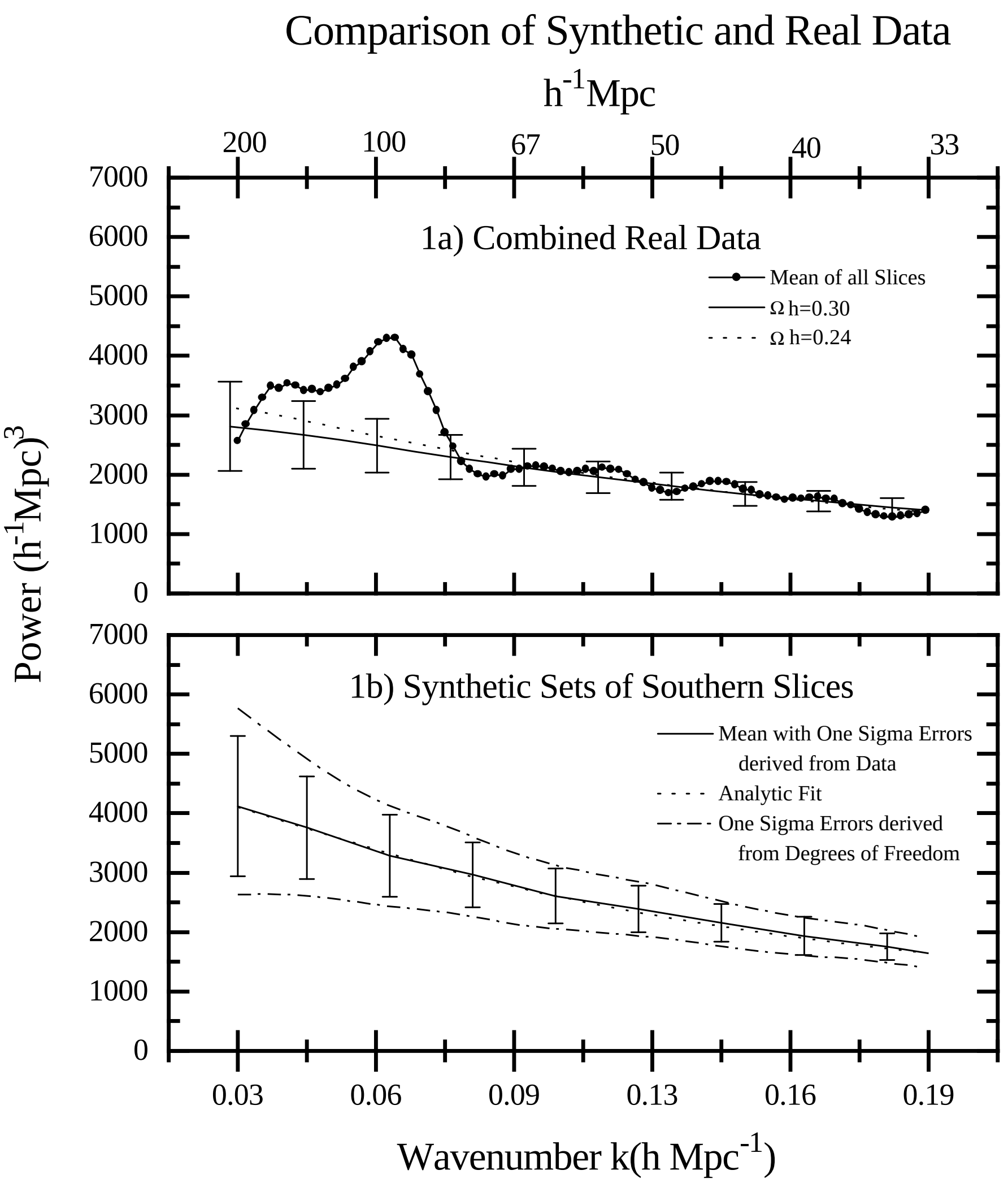
Comparison of Synthetic and Real Data
h-1Mpc
200
100
67
50
40
33
7000
6000
5000
4000
3000
2000
1000
0
1a) Combined Real Data
Mean of all Slices
Ω h=0.30
Ω h=0.24
Power (h-1Mpc)3
1b) Synthetic Sets of Southern Slices
Mean with One Sigma Errors
derived from Data
Analytic Fit
One Sigma Errors derived
from Degrees of Freedom
0.03
0.06
0.09
0.13
0.16
0.19
Wavenumber k(h Mpc-1)

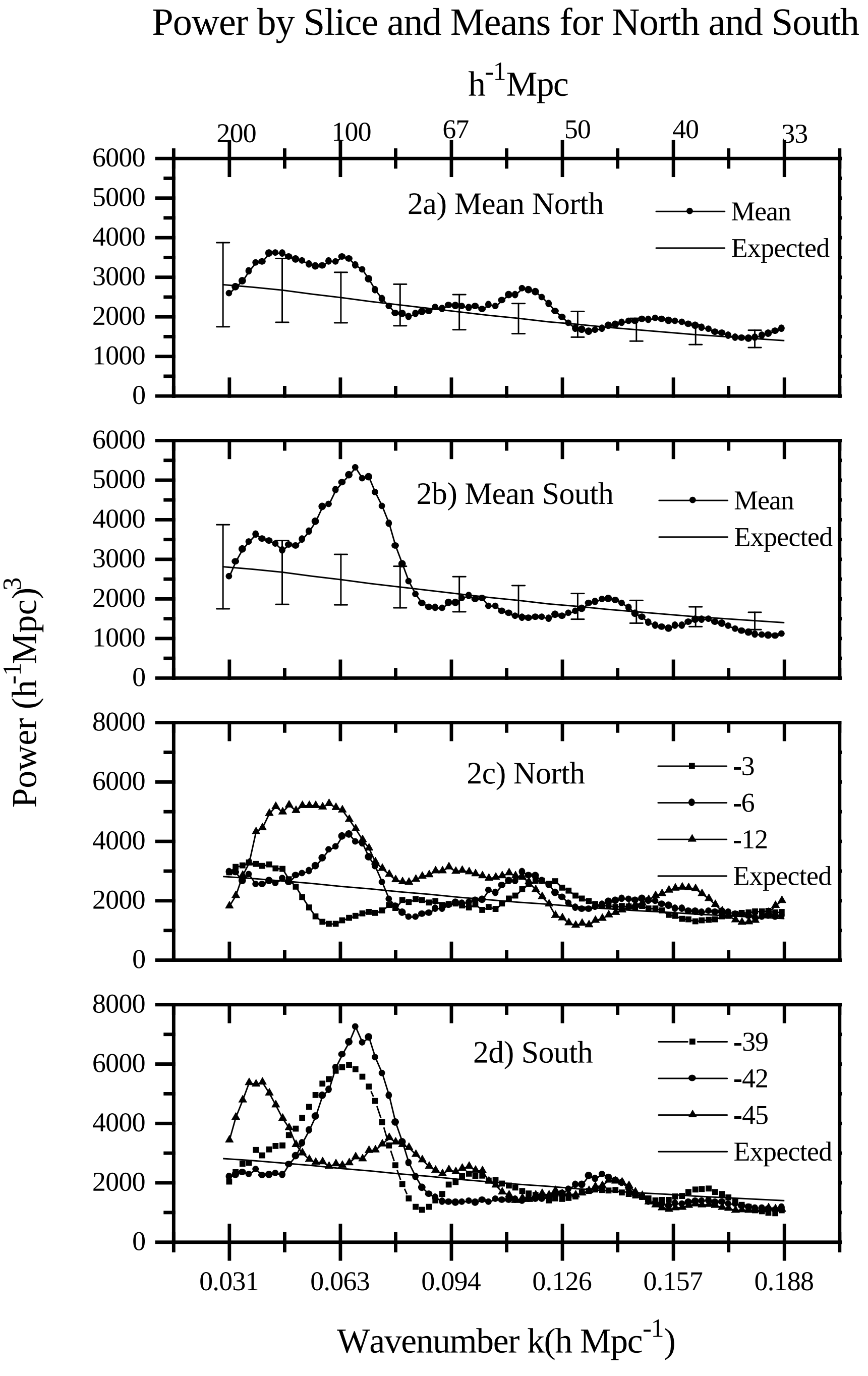
Power by Slice and Means for North and South
h-1Mpc
200
100
67
50
40
33
2a) Mean North
Mean
Expected
2b) Mean South
Mean
Expected
2c) North
-3
-6
-12
Expected
2d) South
-39
-42
-45
Expected
Power (h-1Mpc)3
0.031
0.063
0.094
0.126
0.157
0.188
Wavenumber k(h Mpc-1)

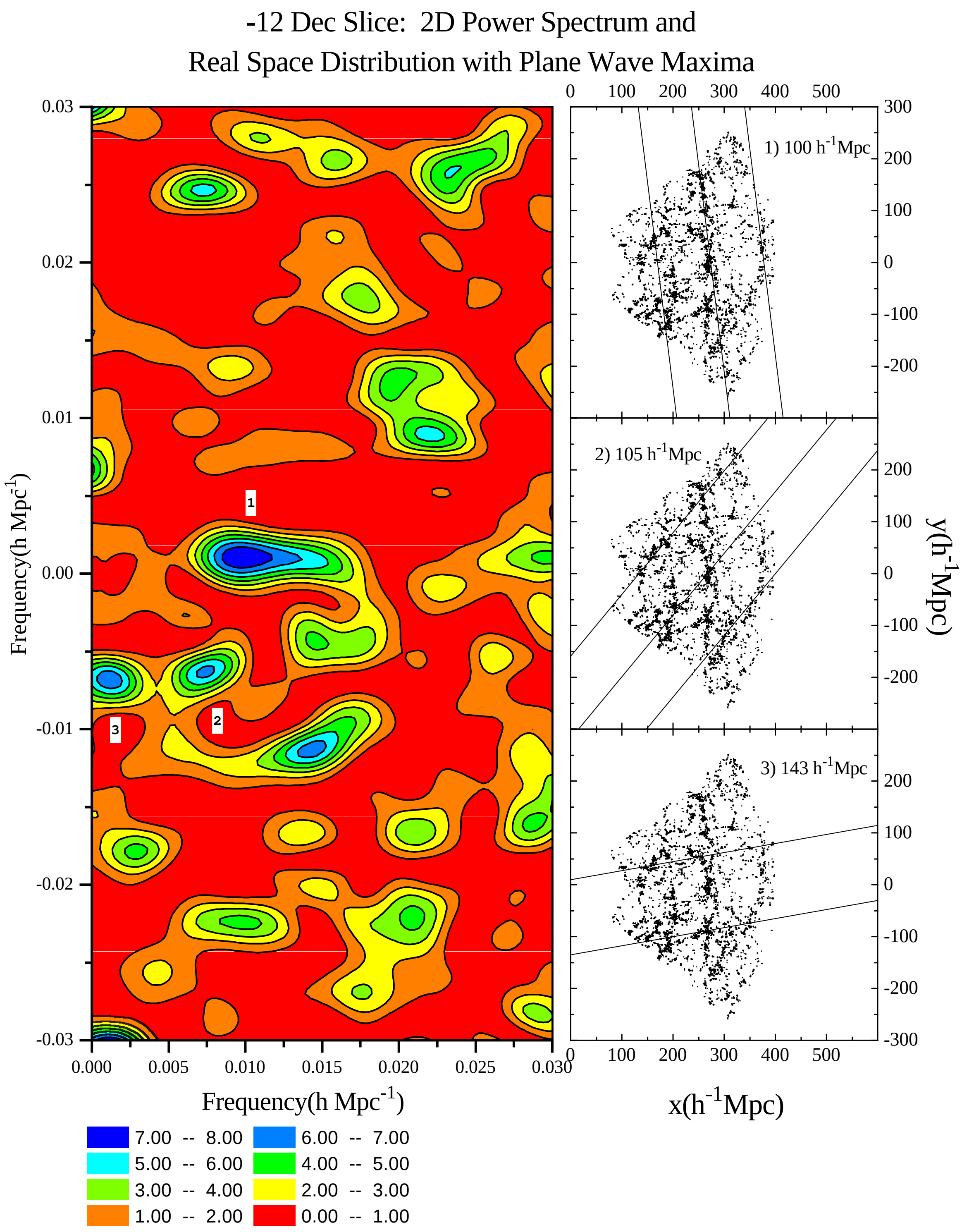

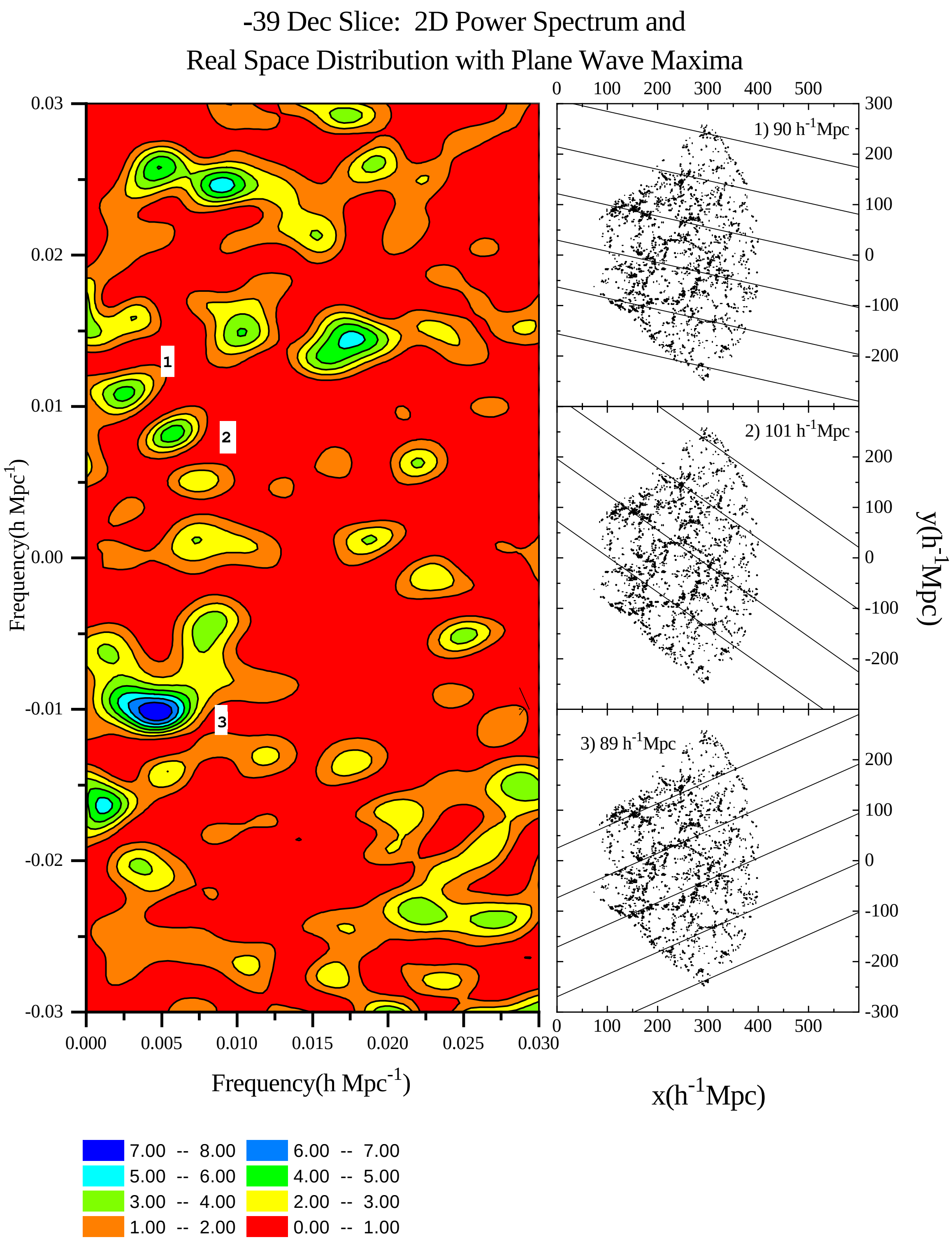